\documentclass[aps,prd,twocolumn,superscriptaddress,showpacs,showkeys]{revtex4}
\usepackage{epsfig}
\usepackage{graphicx}
\usepackage{amsmath,amssymb,amsfonts,latexsym}
\usepackage{graphicx}

\usepackage{epsfig,amssymb,amsfonts,verbatim}

\usepackage{amsmath}
\usepackage{latexsym}
\usepackage{amsfonts}
\usepackage{amssymb}
\usepackage{color}

\Large
\begin{document}

\title{Origin of strange metallic phase in cuprate superconductors}

\author{Andrew Das Arulsamy}

\affiliation{Condensed Matter Group, Division of Interdisciplinary Science, F-02-08 Ketumbar Hill, Jalan Ketumbar, 56100 Kuala-Lumpur, Malaysia}

\date{\today}

\begin{abstract}
The origin of strange metallic phase is shown to exist due to these two conditions---(i) the electrons are strongly interacting such that there are no band and Mott-Hubbard gaps, and (ii) the electronic energy levels are crossed in such a way that there is an electronic energy gap between two energy levels associated to two different wave functions. The theory is also exploited to explain (i) the upward- and downward-shifts in the $T$-linear resistivity curves, and (ii) the spectral weight transfer observed in the soft X-ray absorption spectroscopic measurements of the La-Sr-Cu-O Mott insulator.
\end{abstract}


\maketitle

\section{Introduction}

Zero electrical resistance in conventional superconductors is an impressive quantum mechanical effect first discovered in 1911 by Kamerlingh Onnes~\cite{kamer}. Rightly so, the physics behind this effect also turned out to be somewhat complicated~\cite{bardeen}. Subsequently, the discovery of cuprate superconductors~\cite{bedABS} was a shock because such a discovery is not possible on the basis of Bardeen-Cooper-Schrieffer theory~\cite{bardeen} where the critical temperatures ($T_{\rm sc}$) for these cuprates are relatively ``high'', about 93 K for a YBa$_2$Cu$_3$O$_7$ material~\cite{wu}. The physical notion relevant to high-temperature superconductivity is equally complicated and remains mysterious~\cite{mann}. But in any case, if high-temperature superconductors are found to be superconducting near room temperatures, then they will revolutionize our current technologies beyond recognition, which can also lead to the much-needed eco-friendly technologies. To predict such materials, we need a proper and consistent theory, and for now, we have two main competing theories to explain the origin of high temperature superconductivity---one is based on the spin-fluctuation superglue~\cite{pines}, while the other relies on the resonating valence bonds~\cite{bask2,bask3,bask,bask4,andersonABS}. These theories should predict and explain why $T_{\rm sc}$ can be high, and why there is a pseudogap phase at a certain transition temperature ($T^*$). Further details and excellent reviews on these competing and other exotic theories can be found here~\cite{refe}.

In this work, we will not propose a new mechanism for high-temperature superconductivity, nor any new explanations as to why $T_{\rm sc}$ in cuprates are high, or why $T^*$ exists. Our primary objective is to investigate the normal state resistivity of cuprates above $T_{\rm sc}$, and above $T^*$ if $T_{\rm sc} < T^*$. In particular, we focus on the origin of strange metallic phase using the ionization energy theory (IET) and the Hubbard model in the atomic limit where IET is shown to provide additional features required to understand the strange metals microscopically, such that the strange metallic phase does not exclusively belongs to cuprates. Here, we will need to ignore the origin of (i) superconductivity, and (ii) the pseudogap phase because our theory lacks the specific interaction terms required to obtain the superconductivity and pseudogap effects. In fact, we do not attempt to address the origin of high-temperature superconductivity. In contrast, we research on these two things---(a) the origin of strange metallic phase above $T_{\rm sc}$ such that the ``strangeness'' is shown to be independent of the $T$-linear resistivity and (b) the upward and downward shifts of resistivity versus temperature curves with doping. Even though the second point has been explained earlier within IET (for details, see Refs.~\cite{andPC1,andPLA1,andPSSB,andNOVA1,andNOVA2,jsnm}), but the major ingredient (the crossed energy levels giving rise to non-zero energy level spacing, $\xi \neq 0$) is still missing, which will be incorporated in this work. Obviously, the ``strangeness'' here refers to systematic changes to the $T$-linear resistivity for different doping, $x$ (not temperature, $T$). For example, the whole curve, $\rho(T,x)$ either shifts upward or downward with respect to $x$ in La$_{1-x}$Sr$_x$CuO$_4$. Such shifts are supposedly strange for any metals within the Fermi liquid theory because the changes to the electron-electron ($e$-$e$) and $e$-phonon ($e$-$ph$) scattering do not shift the resistivity curves systematically with respect to the types of chemical elements or $\xi$. We will explain why this strangeness occurs in cuprates and why it is not unique to cuprates such that other strongly correlated matter can exhibit such shifts, including the $T$-linear resistivity effect. 

In the following sections, we will first introduce the notion of energy-level spacing ($\xi$) within the ionization energy theory (IET), including the details on how to approximate the values for $\xi$ in a given system, and for systems with different elemental compositions in the presence of, and in the absence of energy-level crossing. Subsequently, we will rewrite the standard one-band Hubbard model within IET, and derive the relevant Green function as a function of $\xi$. We will then discuss the physical implications of this new Green function in the presence of, and in the absence of energy level crossing, Mott-Hubbard and band gaps. We will then go on to explain why and how the non-zero $\xi$ in the presence of energy-level crossing (zero Mott-Hubbard and band gaps) can be postulated to give rise to a strange metallic phase. To make contact with the experimental data, we make use of the resistivity curves for different elemental compositions, in which, we will show that systematic changes to $\xi$ nicely correspond to the upward or(and) downward shift(s) in the resistivity curves. 

Moreover, the changes in $\xi$ are also used to capture the red- and blue-shift, and the changes in the X-ray absorption spectra peak intensities measured for the La-Sr-Cu-O system without any inconsistency and/or self-contradiction. For example, we will exploit our knowledge on $\xi$ to show that the origin of strange metallic phase in cuprate superconductors is due to crossed energy levels such that the band and Mott-Hubbard gaps are zero, but $\xi \neq 0$. Additional information are also provided based on the proof developed by Parameswaran, Shankar and Sondhi~\cite{shankABS}, which can be used to generalize $\xi$ such that $\xi$ = 0 and $\xi$ = irrelevant-constant imply free-electron and Fermi-liquid metals, respectively. 

\section{Theoretical details}  

In the introduction, we have conveniently denoted the variable $\xi$ as the real (means unique and true) energy level spacing, which simply means, the energy spacing between real energy levels. We now define $\xi$ properly.  

\subsection{Energy-level spacing, $\xi$}  

To technically understand $\xi$, we start from the IET-Schr$\ddot{\rm o}$dinger equation, 
\begin {eqnarray}
&&i\hbar\frac{\partial \Psi(\textbf{r},t)}{\partial t} = \bigg[-\frac{\hbar^2}{2m}\nabla^2 + V_{\rm IET}\bigg]\Psi(\textbf{r},t) \nonumber \\&& = H_{\rm IET}\Psi(\textbf{r},t) = (E_0 \pm \xi)\Psi(\textbf{r},t), \label{eq:IN14}
\end {eqnarray}  
which has been proven to be exact for atoms~\cite{andPRA} where $\Psi(\textbf{r},t)$ is the time-dependent many-body wave function, $\hbar = h/2\pi$, $h$ denotes Planck constant and $m$ is the mass of electron. Here $E_0 \pm \xi$ is the real (true and unique) energy levels (or the total energy if all the occupied levels are summed over) for a given quantum system (atom or molecule or solid). Even though Eq.~(\ref{eq:IN14}) is exact, but it does not have specific interaction terms to address a high-temperature superconductor or any specific system. For example, to study superconductivity and the pseudogap phenomena in cuprates, we need to find and introduce the appropriate interaction terms into Eq.~(\ref{eq:IN14}).

From Eq.~(\ref{eq:IN14}), $\xi$ can be related to every observable, provided that these observables can be evaluated from their corresponding renormalized operators. For example, we can renormalize the screening potential operator to study the screening strength as a function of $\xi$. Anyway, for free-electron systems, Eq.~(\ref{eq:IN14}) non-trivially reduces to the standard Schr$\ddot{\rm o}$dinger equation because $(E_0 \pm \xi) \rightarrow E$. Hence, Eq.~(\ref{eq:IN14}) is valid for systems with (i) non-degenerate energy levels or (ii) energy-level crossings (including degenerate energy levels)~\cite{andPRA}. Here, $E_0$ is the energy levels at zero temperature and in the absence of any external disturbances. Moreover, $\xi$ is known as the ionization energy or the energy-level spacing (proven elsewhere~\cite{iet}) where $+\xi$ is for electrons while $-\xi$ is for holes. 

Now, we can define the renormalized electron self-energy ($\tilde{\Sigma}$) using the renormalized screened Coulomb potential embedded in $V_\textbf{\rm IET}$, which is given by~\cite{andPRA}
\begin {eqnarray}
V_{\rm IET} &=& V^{\rm ion}_{\rm electron} + \tilde{V}_{\rm sc}(\sigma_{\rm IET}) \nonumber \\&=& -\frac{e^2}{4\pi\epsilon_0}\sum_{i,j,I}\bigg[\frac{Z}{|\textbf{r}_i - \textbf{R}_I|} - \frac{1}{|\textbf{r}_i-\textbf{r}_j|}e^{-\sigma_{\rm IET}(r_i + r_j)}\bigg], \nonumber \\&& \label{eq:IN16}
\end {eqnarray}  
where $Z$ is the atomic number, $e$ is the charge of electron, and $\epsilon_0$ denotes the permittivity of free space. Here, $\tilde{V}_{\rm sc}$ denotes the renormalized screened Coulomb potential, giving rise to
\begin {eqnarray}
&&\tilde{\Sigma}(\xi) \nonumber \\&& = -\frac{e^2}{4\pi\epsilon_0}\bigg\langle\Psi(\textbf{r}_{i,j})\bigg|\sum_{i,j}\frac{1}{|\textbf{r}_i-\textbf{r}_j|}e^{-\sigma_{\rm IET}(r_i + r_j)}\bigg|\Psi(\textbf{r}_{i,j})\bigg\rangle, \nonumber \\&& \label{eq:IN16b}
\end {eqnarray}  
where
\begin {eqnarray}
\sigma_{\rm IET} = \mu\exp{\bigg[-\frac{1}{2}\lambda\xi\bigg]}, \label{eq:IN17}
\end {eqnarray}  
$\mu$ is the constant of proportionality, $\lambda = (12\pi\epsilon_0/e^2)a_{\rm B}$, and $a_{\rm B}$ is the Bohr radius~\cite{andPLA2} and $\Psi(\textbf{r}_{i,j})$ is the time-independent many-body wave function. Observe that $\tilde{\Sigma}(\xi) \propto \xi$, and each renormalized variable carries a tilde. The above renormalization procedure is exact and is based on the energy-level spacing renormalization group method~\cite{andAOP}, which can be related (exactly) to the Shankar renormalization technique~\cite{shank1,yale,shank3}. The term, $\tilde{V}_{\rm sc}(\sigma_{\rm IET})$ in Eq.~(\ref{eq:IN16}) and in Eq.~(\ref{eq:IN16b}) has been renormalized (see Eqs.~(2.1) and~(2.2) in Ref.~\cite{andAOP}). In addition, $\tilde{V}_{\rm sc}(\sigma_{\rm IET})$ approaches the Thomas-Fermi approximation when $\xi \rightarrow 0$, and when $r_i + r_j$ is replaced by $|\textbf{r}_i - \textbf{r}_j|$, which is suitable for Fermi metals. Here, $\xi$ = 0 means the $i^{\rm th}$ electron can occupy $j^{\rm th}$ energy level, and vice versa without any energy penalty, and therefore, forms a Fermi metal. Consequently, we do not require discussion on the basis of $|\textbf{r}_i - \textbf{r}_j|$.

In the most general sense, $\xi$ is defined to be a real energy level spacing, which can be taken to mean the energy cost that needs to be overcome when an electron from one energy level tries to occupy another energy level. If these energy levels refer to an occupied level (in the valence band) and an empty level (in the conduction band), then $\xi$ refers to the band gap ($E_g$). On the other hand, if the energy levels now refer to two occupied levels, such that if any one of the electron attempts to occupy the other occupied energy level, then $\xi$ refers to the Mott-Hubbard gap ($\texttt{U}_{\texttt{H}}$). Here, both $E_g$ and $\texttt{U}_{\texttt{H}}$ deal with energy levels residing in two energetically isolated bands where (i) the valence and conduction bands are separated by an energy gap, $E_g$ and (ii) the upper and lower Mott-Hubbard bands are separated by an energy gap of a different kind, $\texttt{U}_{\texttt{H}}$. Moreover, there is not a single energy level that connects the two bands---between the valence and conduction bands or between the lower and upper Mott-Hubbard bands. Therefore, $\xi$ is a generalized gap such that the band ($E_g$) and the Mott-Hubbard ($\texttt{U}_{\texttt{H}}$) gaps are special cases. 

If we allow the two energetically isolated bands to overlap by means of proper alloying, then one can obtain one of these three metals (due to some infinite or finite numbers of energy level crossings), namely, (i) non-interacting Fermi metal (Fermi gas (with infinite numbers of energy level crossings), $\xi = 0, E_g = 0, \texttt{U}_{\texttt{H}} = 0$), (ii) weakly interacting Fermi metal (Fermi liquid, $\xi = {\rm irrelevant~constant}, E_g = 0, \texttt{U}_{\texttt{H}} = 0$), and (iii) strongly interacting metal (strange metal, $\xi \neq 0, E_g = 0, \texttt{U}_{\texttt{H}} = 0$). We attribute the origin of a strange metal to $\xi$, which is neither a zero nor an irrelevant constant. This means that the electron conduction in a strange metal still requires the electron-flow between crossed energy levels, such that a conduction electron needs to overcome the gap introduced by $\xi$ at the energy-level crossing points. 

The conditions $\texttt{U}_{\rm{H}} = 0 = E_g$ and $\xi \neq 0$ can be understood by noting that an energy-level crossing at a certain $\textbf{k}$-point, say at $\textbf{k}_1$ may imply $E_a(\textbf{k}_1) = E_b(\textbf{k}_1)$ but $\xi \neq 0$ due to interactions where $E_a$ and $E_b$ are the eigenvalues for the solved two-level Hamiltonian
\begin {eqnarray}
H(\textbf{k})\varphi_a(\textbf{k}) &=& [H_0(\textbf{k}) + \mathcal{V}(\textbf{k})]\varphi_a(\textbf{k}) \nonumber \\&=& [h_a(\textbf{k}) + \texttt{v}_a(\textbf{k})]\varphi_a(\textbf{k}) = E_a(\textbf{k})\varphi_a(\textbf{k}), \nonumber \\ \\
H(\textbf{k})\varphi_b(\textbf{k}) &=& [H_0(\textbf{k}) + \mathcal{V}(\textbf{k})]\varphi_b(\textbf{k}) \nonumber \\&=& [h_b(\textbf{k}) + \texttt{v}_b(\textbf{k})]\varphi_b(\textbf{k}) = E_b(\textbf{k})\varphi_b(\textbf{k}). \label{eq:xxx5} 
\end {eqnarray}
Here, $H_0(\textbf{k})$ is the non-interacting Hamiltonian, whereas $\mathcal{V}(\textbf{k})$ contains all the interaction terms, and $H(\textbf{k})$ does not have to be a mean field operator. Obviously we have $E_a(\textbf{k})$, $h_a(\textbf{k})$, $\texttt{v}_a(\textbf{k})$, $h_b(\textbf{k})$, $\texttt{v}_b(\textbf{k})$ and $E_b(\textbf{k})$ as eigenvalues. Degeneracy due to energy level crossing in $\textbf{k}$-space implies $E_a(\textbf{k}_1) = E_b(\textbf{k}_1)$ at a certain $\textbf{k}$ point ($\textbf{k}_1$), and at other $\textbf{k}$ points, they are not degenerate. If they are always degenerate, then $h_a(\textbf{k}) = h_b(\textbf{k})$, $\texttt{v}_a(\textbf{k}) = \texttt{v}_b(\textbf{k})$ and $E_a(\textbf{k}) = E_b(\textbf{k})$, which physically mean $\varphi_a(\textbf{k}) = \varphi_b(\textbf{k})$: this strictly implies that there are two electrons occupying the degenerate energy level throughout the $\textbf{k}$-space ($\texttt{U}_{\rm{H}} = 0$ and $\xi = 0$). As a consequence, degeneracy due to energy level crossing requires $h_a(\textbf{k}_1) \neq h_b(\textbf{k}_1)$, $\texttt{v}_a(\textbf{k}_1) \neq \texttt{v}_b(\textbf{k}_1)$ and $E_a(\textbf{k}_1) = E_b(\textbf{k}_1)$. In other words, $\texttt{U}_{\rm{H}} = 0$ (because $E_a(\textbf{k}_1) = E_b(\textbf{k}_1)$) and $\xi \neq 0$ (because $\varphi_a(\textbf{k}) \neq \varphi_b(\textbf{k})$). The above logical exposition originates from Ref.~\cite{andQAT}.

\subsection{One-band IET-Hubbard model and its Green function}

To derive the IET version of the Hubbard model, we start from the one-band Hubbard Hamiltonian~\cite{hub,mott,assa}
\begin {eqnarray}
&&H^{\texttt{H}} = E^{(0)}\sum_{\textbf{R},\sigma}c^{\dag}_{\textbf{R}\sigma}c_{\textbf{R}\sigma} + \texttt{t}\sum_{\textbf{R}+\textbf{d},\sigma}c^{\dag}_{\textbf{R}+\textbf{d}\sigma}c_{\textbf{R}\sigma} \nonumber \\&& + \texttt{U}_{\rm{H}}\sum_{\textbf{R}}n_{\textbf{R}\uparrow}n_{\textbf{R}\downarrow}, \label{eq:1000.16}   
\end {eqnarray}    
where $n_{\textbf{R}\uparrow}$ and $n_{\textbf{R}\downarrow}$ are the number operators with spin-up and spin-down, respectively, $E^{(0)}$ is the non-interacting atomic energy levels, $\textbf{R}$ is the lattice-point coordinate, and $\textbf{d}$ is the distance between two nearest ions with a certain spin configuration. The hopping or transfer matrix elements is denoted by $\texttt{t}$. Here, $\sigma$ can be spin-up or -down where $\sigma$ (or $\sigma'$) and $-\sigma$ (or $-\sigma'$) denote spin-up and -down, respectively. We now use Eqs.~(\ref{eq:IN14}) and~(\ref{eq:IN16b}) to rewrite Eq.~(\ref{eq:1000.16}), and we obtain the one-band IET-Hubbard model Hamiltonian,   
\begin {eqnarray}
H_{\rm IET}^{\texttt{H}} = E_{0}\sum_{\textbf{R},\sigma}c^{\dag}_{\textbf{R}\sigma}c_{\textbf{R}\sigma} + \xi(\sigma,\tilde{\Sigma})\sum_{\textbf{R}}n_{\textbf{R}\uparrow}n_{\textbf{R}\downarrow}. \label{eq:1000.16xx}
\end {eqnarray}    
Fortunately, the condition associated to $\xi$ ($\xi \neq 0$ even if $E_g = 0$ and $\texttt{U}_{\texttt{H}} = 0$) explained earlier using Eq.~(\ref{eq:xxx5}) actually simplify Eq.~(\ref{eq:1000.16}) ``mathematically'' such that $\texttt{t}$ is irrelevant, but the difficulty of solving Eq.~(\ref{eq:1000.16xx}) still remains. The above condition is not a ``simplifying condition'' invoked to simplify Eq.~(\ref{eq:1000.16}). For example, $E_0 \neq E^{(0)}$ where $E_0$ is the real energy levels (or the real bandwidth) of a particular system (in the absence of disturbances at $T = 0$~K), and all the changes to the electron-electron, electron-ion and spin-exchange interactions come as a result of external disturbances, and are taken into account by the self-energy, $\tilde{\Sigma}(\xi) = \xi(\sigma,\tilde{\Sigma})^{-1}$ where $\xi(\sigma,\tilde{\Sigma})^{-1} \neq 1/\xi(\sigma,\tilde{\Sigma})$. Therefore, the complexity remains intact because we still need to determine $E_0$ and $\xi(\sigma,\tilde{\Sigma})$. However, $E_0$ is a constant by definition, and we know exactly how to handle $\xi(\sigma,\tilde{\Sigma})$ by means of the ionization energy approximation, $\xi^{\rm quantum}_{\rm matter} \propto \xi^{\rm constituent}_{\rm atoms}$.

What we did to transform Eq.~(\ref{eq:1000.16}) to Eq.~(\ref{eq:1000.16xx}) was to acknowledge that $\xi(\sigma,\tilde{\Sigma})$ is the real (true and unique) energy level spacing such that 
\begin {eqnarray}
&&\bigg[\texttt{t}\sum_{\textbf{R}+\textbf{d},\sigma}c^{\dag}_{\textbf{R}+\textbf{d}\sigma}c_{\textbf{R}\sigma} + \texttt{U}_{\rm{H}}\sum_{\textbf{R}}n_{\textbf{R}\uparrow}n_{\textbf{R}\downarrow}\bigg] \nonumber \\&& \longrightarrow \xi(\sigma,\tilde{\Sigma})\sum_{\textbf{R}}n_{\textbf{R}\uparrow}n_{\textbf{R}\downarrow}. \label{eq:1000.16xx2}
\end {eqnarray}
This means that, $\xi(\sigma,\tilde{\Sigma})$ is by definition different for each different value from the left-hand side term in Eq.~(\ref{eq:1000.16xx2}) and \textit{vice versa}. For example, to use the original Hubbard model (Eq.~(\ref{eq:1000.16})), one requires to know both $\texttt{t}$ and $\texttt{U}_{\rm{H}}$ in order to determine the hopping probability, in contrast, Eq.~(\ref{eq:1000.16xx}) only requires $\xi(\sigma,\tilde{\Sigma})$ to find the same hopping probability. Later, we will show why and how $\xi^{\rm quantum}_{\rm matter} \propto \xi^{\rm constituent}_{\rm atoms}$ is not always true for solids, for instance, when $\xi^{\rm quantum}_{\rm matter} \rightarrow 0$ (Fermi gas) or when $\xi^{\rm quantum}_{\rm matter} \rightarrow$ irrelevant constant (Fermi liquid). 
     
\begin{figure}[hbtp!]
\begin{center}
\scalebox{0.38}{\includegraphics{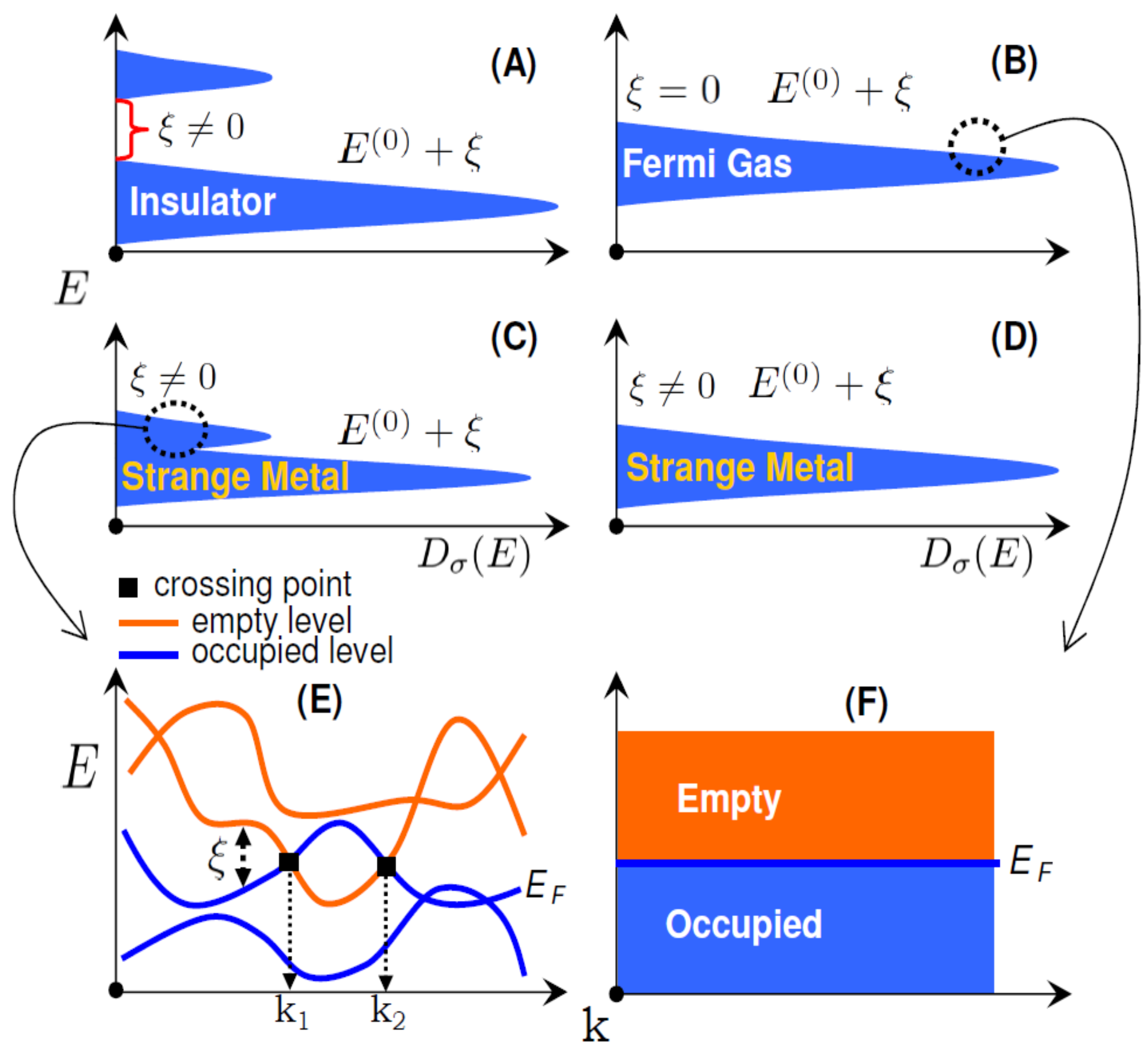}}
\caption{Predicted density of states, $D(\sigma,E)$ based on Eq.~(\ref{eq:1111.21x}). Diagram (A) depicts the usual insulating behavior where $\xi = \texttt{U}_\texttt{H} \neq 0$. In this case, both $\xi$ and $\texttt{U}_\texttt{H}$ represent the Mott-Hubbard gap. Whereas, diagram (B) shows the Fermi gas or free-electron metallic behavior. Here, the upper and lower Mott-Hubbard bands overlap such that $\xi = 0 = \texttt{U}_\texttt{H}$. If $\xi$ is an irrelevant (non-zero) constant and $\texttt{U}_\texttt{H} = 0$, then diagram (B) refers to Fermi liquid. The strange metallic behavior is obtained when the energy levels are crossed in such a way that $\xi \neq \texttt{U}_\texttt{H} = 0$, where $\xi$ remains finite ($\xi \neq 0$), and $\xi$ is not an irrelevant constant, as indicated in diagrams (C) and (D). The diagrams drawn in (E) and (F) indicate the band structure for a strange metal and a Fermi gas, respectively, where $E_F$ is the Fermi level or the chemical potential, which represents the highest occupied energy level. The energy levels in strange metals are crossed (see (E)), and we have indicated two crossing points ($\textbf{k}_1$ and $\textbf{k}_2$) using two $\blacksquare$. For $\textbf{k}$ points other than $\textbf{k}_1$ and $\textbf{k}_2$, $\xi(\textbf{k}) \neq 0$ as shown in (E), for example, $\xi(\textbf{k})$ is the energy level spacing between the highest occupied level ($E_F$) and the lowest empty level. However, one should note that $\xi(\textbf{k})$ is not always zero at the crossing points (see Eq.~(\ref{eq:new100}) and text for details).}   
\label{fig:1}
\end{center}
\end{figure}

The Green function for the one-band IET-Hubbard model (given in Eq.~(\ref{eq:1000.16xx})) can be derived from~\cite{assa,mahan,ziman}, 
\begin {eqnarray}
\mathcal{G}(t-t') = -\frac{i}{\hbar}\theta(t-t')\langle \{c_{\alpha}(t),c^{\dag}_{\alpha}(t')\}\rangle, \label{eq:1000.12}   
\end {eqnarray}    
where, $c_{\alpha}(t)$ and $c^{\dag}_{\alpha}(t)$ denote the usual annihilation and creation operators for fermions, while the curly bracket denotes anti-commutator. The final result is (see the Appendix for a sketched derivation)  
\begin {eqnarray}
\mathcal{G}(E,\xi) = \frac{1-\langle n_{-\sigma}\rangle}{E - E_0 + i\delta} + \frac{\langle n_{-\sigma}\rangle}{E - [E_0 + \xi(\sigma,\tilde{\Sigma})] + i\delta}. \label{eq:1111.21x}   
\end {eqnarray}
Using Eq.~(\ref{eq:1111.21x}) we plot Fig.~\ref{fig:1}, which captures the predicted density of states when $\xi \geq 0$. Here, the condition $\texttt{U}_{\texttt{H}} = 0$ and $\xi \neq 0$ originated from Eq.~(\ref{eq:xxx5}), and this condition is postulated to be responsible for the strange metallic phase in cuprates above $T_{\rm sc}$---provided that $\xi$ is not an irrelevant constant. In Fig.~\ref{fig:1}(E), we can see why $\xi \neq 0$ throughout the momentum space, except at the crossing points ($\blacksquare$). Earlier, we have given the technical reason why $\xi \neq 0$ can exist even at the points where two energy levels cross, namely (see Eq.~(\ref{eq:xxx5})), $h_a(\textbf{k}_1) \neq h_b(\textbf{k}_1)$, $\texttt{v}_a(\textbf{k}_1) \neq \texttt{v}_b(\textbf{k}_1)$ and $E_a(\textbf{k}_1) = E_b(\textbf{k}_1)$, which means 
\begin {eqnarray}
&&\xi(\textbf{k}) = \sum_{i}E_a(\textbf{k}_i) - \sum_{i}E_b(\textbf{k}_i) \neq 0, \nonumber \\&& \sum_{i}E_a(\textbf{k}_i) < \sum_{i}E_b(\textbf{k}_i). \label{eq:new100}   
\end {eqnarray}
Here, $\sum_iE_a(\textbf{k}_i)$ is the highest occupied energy level, while $\sum_iE_b(\textbf{k}_i)$ is the lowest unoccupied energy level, or $\sum_iE_a(\textbf{k}_i)$ is the second highest occupied energy level, whereas, $\sum_iE_b(\textbf{k}_i)$ is the highest occupied energy level. To understand why $\xi \neq 0$ even if $E_a(\textbf{k}_1) = E_b(\textbf{k}_1)$, we need to acknowledge that one requires to overcome an energy cost (see Eq.~(\ref{eq:new100})) due to a wave function transformation, $\varphi_a(\textbf{k}) \longrightarrow \varphi_b(\textbf{k})$ (see Eq.~(\ref{eq:xxx5})). However, note here that this energy cost is not equal to $\xi(\textbf{k})$ given in Eq.~(\ref{eq:new100}), but due to $\sum_{i}E_a(\textbf{k}_i) - \sum_{i}E_b(\textbf{k}_i) \neq 0$, which implies $\varphi_a(\textbf{k}) \neq \varphi_b(\textbf{k})$. In contrast, one does not require such a wave function transformation for Fermi gas metals because $\varphi_a(\textbf{k}) = \varphi_b(\textbf{k})$ (Fig.~\ref{fig:1}(F)) because the energy levels are crossed in such a way that there are no space between crossing points, throughout the momentum space. On the other hand, one can still use Fig.~\ref{fig:1}(E) to represent the Fermi liquid metals, in this case however, $\xi(\textbf{k})$ given in Eq.~(\ref{eq:new100}) is an irrelevant constant. This means that, there is no energy cost that needs to be overcome because both Fermi-liquid and Fermi-gas metals do not require any wave function transformation at the crossing points, and throughout the $\textbf{k}$-space, respectively.        
\section{Strange metallic phase above $T_{\rm sc}$} 

Using the elementary resistivity formula, $\rho = m^*/ne^2\tau_{\rm ee}$ and the carrier density (derived from IET)~\cite{andPC1},
\begin {eqnarray}
n(T,\xi) &=& \int^{\infty}_{0}f(E_0,\xi)D(E_0)dE_0 \nonumber \\&=& \frac{m^*_ek_{\rm B}T}{\pi\hbar^2}\exp{\bigg[\frac{-\xi_{\rm cuprates} + E_{\rm F}^{(0)}}{k_{\rm B}T}\bigg]}, \label{eq:24}   
\end {eqnarray}
one can arrive at 
\begin {eqnarray}
\rho(T,\xi) \propto \frac{1}{n(T,\xi)} \propto \exp{\bigg[\frac{\xi_{\rm cuprates}}{k_{\rm B}T}\bigg]}. \label{eq:24a}   
\end {eqnarray}
where $m^*$ is the electron effective mass, $f(E_0,\xi)$ denotes the ionization energy based Fermi-Dirac statistics (FDS) and $D(E_0) = m^*_e/\pi\hbar^2$, which is the two-dimensional density of states, $k_{\rm B}$ is the Boltzmann constant, and $E_{\rm F}^{(0)}$ is the undisturbed Fermi energy for $T = 0$~K (hence, it is also a constant). In FDS, the chemical potential, $\mu_{\rm ch}$ is for $T$ = 0K, which is equal to $E^{(0)}_{\rm F}$. We have defined $–E^{(0)}_{\rm F}/k_{\rm B}T$ as the temperature-dependent chemical potential. In FDS, when the exponential term (in the denominator) is much larger than one, then FDS approaches the Maxwell-Boltzmann statistics (MBS). In such cases, both FDS and MBS happen to give the same probability, but this does not imply that the electrons are now classical particles obeying MBS. Moreover, $\xi$ cannot exist in MBS.

For a given composition, $\xi_\textbf{\rm cuprates}$ can be taken as a temperature independent constant, but this is not always the case. For example, from the ionization energy approximation, $\xi_\textbf{\rm cuprates} \propto \xi^{\rm constituent}_{\rm atoms}$, and $\xi^{\rm constituent}_{\rm atoms}$ can be determined from~\cite{andPC1,andPLA1,andPSSB} 
\begin {eqnarray}
\xi^{\rm constituent}_{\rm atoms} = \sum_j\sum_i^z \frac{1}{z}\xi_{j,i}(\texttt{X}^{i+}_{j}), \label{eq:24b}   
\end {eqnarray}
where each subscript $j$ represents one type of chemical element ($\texttt{X}_j$) in a particular cuprate, while $i = 1, 2, \cdots, z$, where $i$ counts the number of valence electrons coming from each chemical element. Equation~(\ref{eq:24b}) implies $\xi^{\rm constituent}_{\rm atoms}$ can be $T$-dependent if the valence state of a multi-valent element changes with decreasing temperature. If the valence-state fluctuates due to some external disturbances, then so does $\xi$ (from Eq.~(\ref{eq:24b})). The experimental proofs for this effect (changing valence state with decreasing temperature) were reported by Dionicio~\cite{dio} using the results of Fukuda \textit{et al}.~\cite{fukud} (see Fig.6.5 in Ref.~\cite{dio}). This additional $T$-dependence gives rise to an increasing $\xi$ with decreasing $T$. In the following paragraphs, we apply Eqs.~(\ref{eq:24}),~(\ref{eq:24a}) and~(\ref{eq:24b}) to the resistivity data obtained from La$_{2-x}$Sr$_x$CuO$_4$.

There are two distinct normal state electronic properties that we will tackle here---(a) why and how $\rho(T,x)$ shifts downward with increasing $x$, and (b) the origin of a strange metallic phase. Here, the ``strangeness'' has got nothing to do with $T$-linear resistivity, but it is due to non-zero and relevant $\xi$, and therefore, we will not develop a resistivity formula to fit the resistivity curves. In other words, any association that may exist between a $T$-linear resistivity and a strange metal is just another ``lucky coincidence''. For example, the $T$-linear resistivity originates from the $T$-dependent scattering-rate~\cite{casey,casey2} and carrier density ($n(T)$)~\cite{andPC1,andPLA1}.   

The inequality, $\xi_{\rm Sr^{2+}} < \xi_{\rm La^{3+}}$ is responsible for the downward shift of $\rho(T)$ with increasing $x$ (see Fig.2 in Ref.~\cite{casey}). For example, $\xi_{\rm Sr^{2+}} = 807$ kJmol$^{-1}$ is less than $\xi_{\rm La^{3+}} = 1152$ kJmol$^{-1}$, and therefore, when one substitutes Sr for La, the energy level spacing ($\xi$) for La$_{2-x}$Sr$_x$CuO$_4$ decreases systematically with increasing $x$, hence promotes electron conduction, which shifts the whole $\rho(T,\xi)$ curve downward following Eq.~(\ref{eq:24a}). Here, we used the raw experimental ionization energy values from Ref.~\cite{web}, and have them averaged using Eq.~(\ref{eq:24b}), and we obtained these values, 807 kJmol$^{-1}$ and 1152 kJmol$^{-1}$ for $\xi(\rm Sr^{2+})$ and $\xi(\rm La^{3+})$, respectively. 

Now, the strange metallic phase arises when $\xi_{\rm cuprates} < k_{\rm B}T$. However, we cannot theoretically determine the value for $\xi_{\rm cuprates}$, which is a real (unique and true) quantity of La$_{2-x}$Sr$_x$CuO$_4$, for a given $x$. In short, what we did was to invoke the ionization energy approximation, calculate the averaged constituent atomic ionization energies for the relevant chemical elements (for Sr$^{2+}$ and La$^{3+}$) and substitute these values into Eq.~(\ref{eq:24a}). Consequently, we obtained the correct resistivity shift with $x$ (see Fig.2 in Ref.~\cite{casey}), and the origin of the metallic phase has been postulated to be due to $\xi \neq 0$ and $\xi$ is a relevant constant.

\subsection{Fermi gas and Fermi liquid are special cases}

Here, we justify that the Fermi gas and Fermi liquid are special cases within the ionization energy theory by exploiting the proof developed by the trio, Parameswaran, Shankar and Sondhi with respect to the bare coupling constant ($g_0$) such that the Fermi gas exists in the limit $g_0 \rightarrow 0$, while $g_0 > 0$ refers to Fermi liquid~\cite{shankABS}. The relevant result is the two-point Cooper-pair correlation function~\cite{shankABS},  
\begin {eqnarray}
&&\Gamma(\bar{\Omega};g_0,\Lambda_{\rm Shankar}) = \frac{1}{2\pi \textsc{v}_{\rm F}}\bigg[\frac{a|\log{(\bar{\Omega}/\Lambda_{\rm Shankar})}|}{1 + ag_0|\log{(\bar{\Omega}/\Lambda_{\rm Shankar})}|}\bigg], \nonumber \\&& \label{eq:25}   
\end {eqnarray}
where $\Omega$ is the $s$-wave Cooper-pair frequency, $\bar{\Omega} = (\Omega^2 + \textsc{P}^2)^{1/2}$, $\textsc{P}$ is the $s$-wave Cooper-pair momentum, $a$ denotes a positive constant, $\Lambda_{\rm Shankar}$ is the Shankar cutoff parameter, $\textsc{v}_{\rm F}$ denotes the Fermi velocity, and Eq.~(\ref{eq:25}) must satisfy $\bar{\Omega} < \Lambda_{\rm Shankar}$. For Fermi gas (free or non-interacting electrons), $g_0 = 0$ and for $T = 0$~K, one has $\Gamma(\bar{\Omega};g_0,\Lambda_{\rm Shankar}) \propto |\log{(\bar{\Omega}/\Lambda_{\rm Shankar})}|$, which diverges logarithmically as $\bar{\Omega} \rightarrow 0$. On the contrary, for Fermi liquid ($g_0 > 0$), $\Gamma(\bar{\Omega};g_0,\Lambda_{\rm Shankar}) \rightarrow 1/2\pi \textsc{v}_{\rm F}g_0$, which converges to a constant as $\bar{\Omega} \rightarrow 0$. This result clearly indicate that Fermi gas is indeed a special case within the Fermi liquid formalism, which can also be associated to IET. In particular, if $\xi$ ceases to be relevant, one can take $\xi = 0$ because $\xi$ is now an irrelevant constant, giving rise to Fermi gas, similar to $g_0 = 0$. Obviously, there is a one-to-one correspondence between the Fermi liquid and Fermi gas when $\xi$ transforms from being an irrelevant (non-zero) constant (Fermi liquid) to zero (Fermi gas). This result is also in agreement with our earlier claim~\cite{andAOP} that a Fermi gas metal (or a free-electron metal) has got to be an emergent phase. For example, we can write~\cite{andAOP} $\Lambda_{\rm Shankar} = (\hbar k^2/2m^*)\Lambda_{\rm IET}$ and replacing $\Lambda_{\rm Shankar}$ with $(\hbar k^2/2m^*)\Lambda_{\rm IET}$ in Eq.~(\ref{eq:25}) does not change the above divergence or convergence in any way. In particular, for the Fermi liquid systems, $\Lambda_{\rm IET} = \exp{(\lambda\xi)}$ is a constant because $\xi$ is itself a constant, while $\lambda = (12\pi\epsilon_0/e^2)a_{\rm B}$ is just a collection of constants defined earlier. As for the Fermi gas, $\Lambda_{\rm IET} \rightarrow 1$ as $\xi \rightarrow 0$. Therefore, replacing $\Lambda_{\rm Shankar}$ with $(\hbar k^2/2m^*)\Lambda_{\rm IET}$ within a Fermi gas or a Fermi liquid system does not change anything physically, as it should be~\cite{andPRA,andAOP}, except that the Fermi gas is a special case exists when and only when $\xi = 0$. On the other hand, Fermi liquid is a special case exists when $\xi$ is an irrelevant (non-zero) constant.   

\section{Further analysis: Soft X-ray absorption spectra}

Contrary to our models derived earlier~\cite{physB1,physB2,physC2}, the particles responsible for the strange metallic behavior discussed in this work are still electrons, but with strong correlation such that it requires $\xi \neq 0$ and $\xi$ is relevant (not an irrelevant constant). Free-electron metals simply require $\xi = 0$, while Fermi liquid embeds the condition, $\xi \neq 0$ but its magnitude is an irrelevant constant. Therefore, our strange metal is due to strongly correlated electrons requiring $\xi \neq 0$, where $\xi$ is the energy that an electron has to pay when it hops from one energy-level to another. These energy levels are crossed in the strange metallic phase.

In this section, we will demonstrate why and how $\xi$ captures the hopping of particles with different elements and elemental composition, without violating the particle-hole symmetry, without any \textit{ad hoc} physical interpretation. Our analyses follow Ref.~\cite{nano}. We show Eq.~(\ref{eq:1111.21x}) correctly captures the spectral weight transfer mechanism in accordance with the data obtained from the soft X-ray absorption spectra, measured by Chen \textit{et al}.~\cite{chen}. These spectra correspond to the changes in the concentrations of Sr$^{2+}$, La$^{3+}$, Cu$^{2+}$, Cu$^{3+}$ and O$^{2-}$ in the compound, La$^{3+}_{2-x}$Sr$^{2+}_x$Cu$^{2+}_{1-y}$Cu$^{3+}_{y}$O$^{2-}_{4+\delta}$. Substituting Sr$^{2+}$ into La$^{2+}_2$Cu$^{2+}_{1-y}$Cu$^{3+}_{y}$O$^{2-}_{4.005}$ will give rise to changing $\delta$ and $y$, which have been discussed in detail elsewhere for other oxides and multi-element solids~\cite{andPSSB,jsnm,cpc,mahen,zung}. For example, such changes can be due to defects and the type of such defects. In particular, since $\delta = 0.005$, we need $y = 0.01/3$ to preserve stoichiometry. In the presence of Sr ($0 \leq x \leq 0.15$), both $y$ and $\delta$ can change, satisfying the linear equation, 
\begin {eqnarray}
x - y - 8 = 2\delta, \label{eq:26}   
\end {eqnarray}
where $\{x,y\} \in [0,1]$ and $\delta \in [0,\frac{1}{2}]$. 

\begin{figure}[hbtp!]
\begin{center}
\scalebox{0.35}{\includegraphics{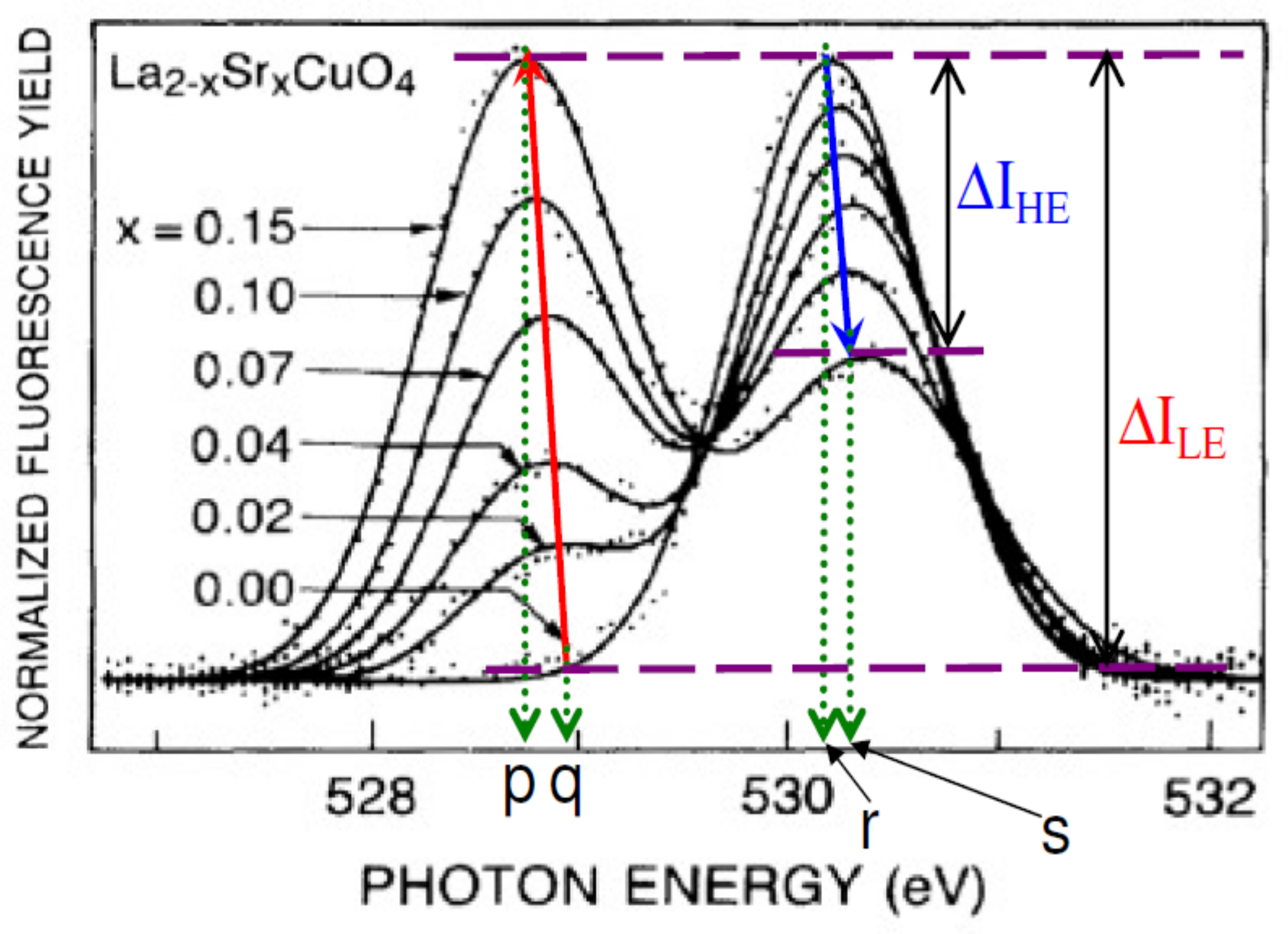}}
\caption{Soft X-ray absorption spectra at the oxygen-K edge for the sample, La$^{3+}_{2-x}$Sr$^{2+}_x$Cu$^{2+}_{1-y}$Cu$^{3+}_{y}$O$^{2-}_{4+\delta}$ obtained by Chen \textit{et al}.~\cite{chen}. The main figure with experimental data points was carbon-copied from Ref.~\cite{chen}. The intensities ($I$) for these peaks have been normalized, backgrounds subtracted, and the data were fitted with a Gaussian function. There are two main peaks, one being the high-energy (HE) peak (occurring between 530 and 531 eV), while the one to the left is the low-energy (LE) peak occurring slightly below 529 eV. The photon energies labeled with $p$ and $q$ show the red-shift (from $q$ to $p$) in the LE peak for increasing $x$, while the blue-shift occurs from $r$ to $s$ in the HE peak for the same doping concentration ($0 \leq x \leq 0.15$). These shifts satisfy the inequality, $p < q < r < s$ where $q - p = \Delta^{\rm red}_{\rm shift}$, $s - r = \Delta^{\rm blue}_{\rm shift}$, $r - q = \Delta_{\rm min}$, and $s - p = \Delta_{\rm max}$. Here, $\Delta^{\rm red}_{\rm shift}$ and $\Delta^{\rm blue}_{\rm shift}$ denote the total red and blue shifts for the stated doping levels, while $\Delta_{\rm min}$ and $\Delta_{\rm max}$ are the minimum and maximum changes to the electrons energy levels, again within the investigated doping range. The change in the normalized intensities for the high- and low-energy sectors are denoted by $\Delta I_{\rm HE}$ and $\Delta I_{\rm LE}$, respectively.}  
\label{fig:2}
\end{center}
\end{figure}

There are four interesting features in the spectra plotted in Fig.~\ref{fig:2}. The first being $p$ and $q$ for $x > 0$ appear at energies lower than $r$ and $s$ where $I_{\rm LE} = 0$ if $x = 0$ (undoped sample). Second, the normalized intensities, $I_{\rm LE}$ increases, while $I_{\rm HE}$ decreases with $x$. Third, the red ($\Delta^{\rm red}_{\rm shift}$) and blue ($\Delta^{\rm blue}_{\rm shift}$) shifts appear for the low- and high-energy peaks, respectively, with increasing $x$, and finally there are some asymmetric changes to the normalized intensities for the range $0 \leq x \leq 0.15$, such that $\Delta I_{\rm LE} > \Delta I_{\rm HE}$ (see Fig.~\ref{fig:2}). In other words, $I_{\rm LE}$ increases faster than $|{\rm d}I_{\rm HE}/{\rm d}x|$. Recall here that the doping concentration, $x$ refers to Sr$^{2+}$ content, which means $x \propto \xi(\rm Sr^{2+})$, $y \propto \xi(\rm Cu^{3+})$, and the normalized carrier density, $\langle n_{-\sigma}\rangle \propto n(T,\xi)$ in the IET Green function (see Eqs.~(\ref{eq:24}) and~(\ref{eq:1111.21x})) is inversely proportional to $\xi$. This means that, increasing $x$ increases $I_{\rm LE}$, while reduces $I_{\rm HE}$, which give rise to an increasing or decreasing carrier density ($\langle n_{-\sigma}\rangle$) with increasing $x$ or increasing $y$, respectively. Here, $\langle n_{-\sigma}\rangle$ decreases with increasing $y$ because $\xi(\rm Cu^{3+}) > \xi(\rm Sr^{2+})$. 

The answer to the first feature is straightforward, it is due to the fact that $\xi_{\rm Sr^{2+}}$ (807 kJmol$^{-1}$) $<$ $\xi_{\rm La^{3+}}$ (1152 kJmol$^{-1}$), which is also in agreement with $\rho$ measurements presented earlier. This means that the electrons with low ionization energies (from Sr$^{2+}$) need low photon energies to be excited, compared to the electrons with large ionization energies (from La$^{3+}$). The change in the normalized intensities are due to increasing $x$ and decreasing $2-x$ for the low- and high-energy sectors, respectively. The red and blue shifts in the spectral weight transfer are also depicted in Fig.~\ref{fig:2} with appropriate arrows. The red-shift originates from the concentration of Sr$^{2+}$, or due to increasing $x$. This implies that more electrons from Sr$^{2+}$ are needed to establish this LE peak at the lowest possible energy as a result of $\xi_{\rm Sr^{2+}}$ $<$ $\xi_{\rm La^{3+}}$. 

Recall that IET requires electrons with the lowest atomic ionization energies to form the valence electrons in a non-Fermi gas compound containing that particular atom. These (valence) electrons also interact weakly with the core electrons (coming from La and Cu), giving rise to smaller energy level spacings. The core electrons form large energy level spacings due to strong interaction among the core electrons~\cite{andPRA}. As a consequence, the HE peak corresponds to the core electrons (due to large ionization energy values) coming from La$^{3+}_{2 - x}$. These core electrons interact more strongly with other core electrons coming from Cu$^{2+}_{1-y}$ and Cu$^{3+}_y$ ($\xi_{\rm Cu^{2+}} = 1352$ kJmol$^{-1}$ and $\xi_{\rm Cu^{3+}} = 2086$ kJmol$^{-1}$) giving rise to HE peak. This strong electron-electron interaction also explains why the energy distribution can and will spread from low to high energies as a result of the interaction among the core electrons and between the valence and core electrons.

We know from Eq.~(\ref{eq:26}) that increasing $x$ causes $y$ to increase to maintain stoichiometry (even in the presence of defects), and therefore, the decreasing number of large ionization energy electrons from La$^{3+}_{2-x}$ are systematically being compensated by the large ionization energy electrons from Cu$^{3+}$ (due to increasing $y$). This scenario is further enhanced if $\delta$ is also found to increase with increasing $x$. Therefore, we can now see the reason why HE peak blue shifts and $\Delta I_{\rm LE} > \Delta I_{\rm HE}$; they are due to increasing $y$, $2 - x \geq 1$ and/or increasing $\delta$. For example, in the absence of this compensation effect, the HE peak should red shift with decreasing $2-x$. But this is not the case, as confirmed by the fact that when $y$ increases, the peak will not disappear even when $x = 1$ because $2 - x \neq 0$ and $y \propto x$. Note here that we did not assume La$^{3+}$ contribute three electrons, while each Sr$^{2+}$ gives two electrons. As a matter of fact, the carrier density increases with increasing Sr$^{2+}$ in accordance with Eq.~(\ref{eq:24}). Therefore, the compensation effect slows down $I_{\rm HE}$ from decreasing as fast as the increasing $I_{\rm LE}$, and consequently $\Delta I_{\rm LE} > \Delta I_{\rm HE}$. Indeed, the blue-shift is due to $\xi_{\rm Cu^{2+,3+}}$ (1352, 2086 kJmol$^{-1}$) $>$ $\xi_{\rm La^{3+}}$ (1152 kJmol$^{-1}$), which means, Cu$^{2+}$ also contributes to the HE peak.

One may wonder what will happen to $\xi_{\rm cuprates}$ due to this compensation effect, will it still decrease with $x$? to maintain the downward-shift of the resistivity curves with increasing $x$ (see Fig.2 in Ref.~\cite{casey}). The value for $\xi_{\rm cuprates}$ will decrease as required if and only if (i) $y < 0.0248$ (Cu$^{3+}$ content is less than 2.48$\%$ and Cu$^{2+}$ $>$ 97.52$\%$ for $x = 0.15$), and (ii) $4 + \delta < 4 - 0.0372$ (oxygen vacancies should be more than 0.042 where 0.042 = 0.0372 + 0.005 for $x = 0.15$). These values in the inequalities were obtained from Eq.~(\ref{eq:26}) and the ionization energy averaging. We can use the same linear equation and the averaging to calculate the above inequalities for $y$ and $4+\delta$, for each different $x$. These are our quantitative predictions. For more examples, see Ref.~\cite{andPSSB} and references therein. 

Finally, the above-stated values ($x$, $y$ and $\delta$) can be confirmed experimentally following the chemical technique developed by Mahendiran \textit{et al}.~\cite{mahen}. Moreover, the systematic doping-dependent spectrum depicted in Fig.~\ref{fig:2} indirectly measures the real $\xi_{\rm cuprate}$ such that $\xi$(La$^{3+}_{2-x}$Sr$^{2+}_x$Cu$^{2+}_{1-y}$Cu$^{3+}_{y}$O$^{2-}_{4+\delta}$) $\approx (p + q)/2$ where $\xi$(La$^{3+}_{2-x}$Sr$^{2+}_x$Cu$^{2+}_{1-y}$Cu$^{3+}_{y}$O$^{2-}_{4+\delta}$) is the top most occupied energy level (for example see Refs.~\cite{nano,amanda} for more examples other than cuprates).

\section{Conclusions} 
    
We have shown the existence of a new generalized energy gap, which is related to the energy level spacing, $\xi$ such that (i) $\xi$ refers to a band gap ($E_g$) if the energy level spacing is between an occupied level in the valence band and an empty level in the conduction band, and (ii) $\xi$ refers to the Mott-Hubbard gap ($\texttt{U}_{\texttt{H}}$) if the energy level spacing is between two occupied levels. These two points mean $\xi = E_g$ and $\xi = \texttt{U}_{\texttt{H}}$. However, there is an additional gap between energy levels at the points where they are crossed in momentum space due to different wave functions ($\psi_a(\textbf{k}) \neq \psi_b(\textbf{k})$, $a \neq b$) associated to these energy levels, such that $\xi \neq 0$ even if $E_g = 0$ and $\texttt{U}_{\texttt{H}} = 0$. Therefore, $\xi$ is indeed a generalized gap in which, the band and the Mott-Hubbard gaps are special cases. We have evaluated that these conditions do exist using the one-band IET-Hubbard model Hamiltonian and the Green function formalism.

The strange metallic phase is postulated to obey this special condition, $\xi \neq 0$ (and $\xi$ is a relevant constant), $E_g = 0$, and $\texttt{U}_{\texttt{H}} = 0$. We also have provided sufficient arguments using IET that the strange metallic phase in cuprates is not unique as it is believed to be because the ``strangeness'' is not due to the $T$-linear resistivity. Using the proof developed by Parameswaran, Shankar and Sondhi, we have further justified that $\xi = 0$ (and $E_g = 0$, $\texttt{U}_{\texttt{H}} = 0$) lead us to a Fermi gas (non-interacting electrons). Whereas, a Fermi liquid satisfies the condition, $\xi \neq 0$ and $\xi$ is an irrelevant constant. Therefore, both Fermi gas and Fermi liquid are special cases within the ionization energy theory.

All of our analyses were shown to obey the resistivity measurements, and the soft X-ray spectra obtained from La$_{2-x}$Sr$_x$CuO$_4$ self-consistently. This means that we can readily extend our experimental analyses on La$_{2-x}$Sr$_x$CuO$_4$ to other correlated systems. We also explained why and how the spectral weight transfer evolves when one changes the types of chemical elements and their concentrations. It would be interesting however, if we could find ways to exploit this energy-level spacing such that it can be associated to the origin of superconductivity and the pseudogap phase, which have been excluded systematically for convenience. But this is another story for another day. However, we anticipate (due to the existence of a strange metallic phase as a result of a special energy-level spacing, $\xi \neq 0$) that a Fermi liquid (or a Fermi gas) is a prerequisite for high-$T_{\rm sc}$ superconductivity, somewhat in agreement with Anderson and Casey~\cite{casey,casey2}.  

\section*{Acknowledgment}
This work was supported by Sebastiammal Innasimuthu and Arulsamy Innasimuthu. I am grateful to Amelia Das Anthony, Malcolm Anandraj and Kingston Kisshenraj for their kind and continuous hospitality. Part of this work, including the Appendix, were completed while I visited the Condensed Matter Group, Division of Mathematics and Physics, Glebe, Sydney, Australia, between Aug/2009 and Feb/2010, hosted by the late Madam Kithriammal Soosay. Special thanks to several referees for pointing out the mistakes, the confusing parts and the issues that require explanations.

\appendix
\section{Derivation of Eq.~(\ref{eq:1111.21x})}

We start from Eq.~(\ref{eq:1000.12})
\begin {eqnarray}
&&\mathcal{G}(t-t') \nonumber \\&& = -\frac{i}{\hbar}\theta(t-t')\langle \{c_{\alpha}(t),c^{\dag}_{\alpha}(t')\}\rangle \nonumber \\&& = -\frac{i}{\hbar}\theta(t-t')\langle c_{\alpha}(t)c^{\dag}_{\alpha}(t') + c^{\dag}_{\alpha}(t')c_{\alpha}(t)\rangle \label{eq:A1}.   
\end {eqnarray}    
\begin {eqnarray}
&&\frac{\partial\mathcal{G}(t-t')}{\partial t} \nonumber \\&& = -\frac{i}{\hbar}\frac{\partial \theta(t-t')}{\partial t}\langle c_{\alpha}(t)c^{\dag}_{\alpha}(t')\rangle - \frac{i}{\hbar}\theta(t-t')\bigg\langle \frac{\partial c_{\alpha}(t)}{\partial t}c^{\dag}_{\alpha}(t')\bigg\rangle \nonumber \\&& -\frac{i}{\hbar}\frac{\partial \theta(t-t')}{\partial t}\langle c^{\dag}_{\alpha}(t')c_{\alpha}(t)\rangle - \frac{i}{\hbar}\theta(t-t')\bigg\langle c^{\dag}_{\alpha}(t')\frac{\partial c_{\alpha}(t)}{\partial t}\bigg\rangle. \nonumber \\&& \label{eq:A2}   
\end {eqnarray}    
Using  
\begin {eqnarray}
\theta(t-t') = \int\delta(t-t')dt, \label{eq:1000.12x} 
\end {eqnarray} 
\begin {eqnarray}
&&i\hbar\frac{\partial\mathcal{G}(t-t')}{\partial t} = \delta(t-t')\langle \{c_{\alpha}(t),c^{\dag}_{\alpha}(t')\}\rangle \nonumber \\&& - \frac{i}{\hbar}\theta(t-t')\bigg\langle\bigg\{i\hbar\frac{\partial c_{\alpha}(t)}{\partial t},c^{\dag}_{\alpha}(t')\bigg\}\bigg\rangle \label{eq:A3}.   
\end {eqnarray}    
Next, we use 
\begin {eqnarray}
\langle c_{\alpha}(t)\rangle = \langle\varphi(t)|c_{\alpha}(t_1)|\varphi(t)\rangle, \label{eq:A4}   
\end {eqnarray}    
\begin {eqnarray}
&&i\hbar\frac{\partial\langle c_{\alpha}(t)\rangle}{\partial t} = \bigg(i\hbar\frac{\partial \langle\varphi(t)|}{\partial t}\bigg)c_{\alpha}(t_1)|\varphi(t)\rangle \nonumber \\&& + \langle\varphi(t)|c_{\alpha}(t_1)\bigg(i\hbar\frac{\partial |\varphi(t)\rangle}{\partial t}\bigg) \label{eq:A5}.
\end {eqnarray}    
Subsequently, we invoke $(i\hbar\partial \langle\varphi(t)|)/\partial t = \langle\varphi(t)|(-H)$,
\begin {eqnarray}
&&i\hbar\frac{\partial\langle c_{\alpha}(t)\rangle}{\partial t} = \langle\varphi(t)|(-H)c_{\alpha}(t_1)|\varphi(t)\rangle + \langle\varphi(t)|c_{\alpha}(t_1)H|\varphi(t)\rangle \nonumber \\&& = \langle\varphi(t)|[c_{\alpha}(t_1),H]|\varphi(t)\rangle \label{eq:A6}.
\end {eqnarray}    
The dummy time label $t_1$ is replaced with $t$, using Eq.~(\ref{eq:A6}) and Eq.~(\ref{eq:A3}), 
\begin {eqnarray}
&&i\hbar\frac{\partial\mathcal{G}(t-t')}{\partial t} = 1 - \frac{i}{\hbar}\theta(t-t')\bigg\langle\bigg\{[c_{\alpha}(t),H],c^{\dag}_{\alpha}(t')\bigg\}\bigg\rangle \nonumber \\&& \label{eq:A7}.   
\end {eqnarray}    
We further invoke $H = H^{\texttt{H}}_{\rm IET}$, and $\delta(t-t') = 1 \longrightarrow \{c_{j\sigma}(t),c^{\dag}_{j\sigma}(t')\} = 1$ to evaluate
\begin {eqnarray}
&&[c_{\alpha}(t),H^{\texttt{H}}_{\rm IET}] = c_{i\sigma}\big(E^{(0)}c^{\dag}_{j\sigma'}c_{j\sigma'} + \xi c^{\dag}_{j\sigma'}c_{j\sigma'}c^{\dag}_{j-\sigma'}c_{j-\sigma'}\big) \nonumber \\&& - \big(E^{(0)}c^{\dag}_{j\sigma'}c_{j\sigma'} + \xi c^{\dag}_{j\sigma'}c_{j\sigma'}c^{\dag}_{j-\sigma'}c_{j-\sigma'}\big)c_{i\sigma} \nonumber \\&& = E^{(0)}\big(c_{i\sigma}c^{\dag}_{j\sigma'}c_{j\sigma'} - c^{\dag}_{j\sigma'}c_{j\sigma'}c_{i\sigma}\big) \nonumber \\&& + \xi\big( c_{i\sigma}c^{\dag}_{j\sigma'}c_{j\sigma'}c^{\dag}_{j-\sigma'}c_{j-\sigma'} - c^{\dag}_{j\sigma'}c_{j\sigma'}c^{\dag}_{j-\sigma'}c_{j-\sigma'} c_{i\sigma}\big), \nonumber \\&& \label{eq:A8}   
\end {eqnarray}    
where $i$ and $j$ are dummy indices replacing $\textbf{R}$,   
\begin {eqnarray}
E^{(0)}\big(c_{i\sigma}c^{\dag}_{j\sigma'}c_{j\sigma'} - c^{\dag}_{j\sigma'}c_{j\sigma'}c_{i\sigma}\big) = E^{(0)}c_{i\sigma}. \label{eq:A9}   
\end {eqnarray}  
Next,  
\begin {eqnarray}
&&\xi\big(c_{i\sigma}c^{\dag}_{j\sigma'}c_{j\sigma'}c^{\dag}_{j-\sigma'}c_{j-\sigma'}\big) \nonumber \\&& = \xi\big(c_{j\sigma'}c^{\dag}_{j-\sigma'}c_{j-\sigma'}\delta_{ij}\delta_{\sigma\sigma'} + c^{\dag}_{j\sigma'}c_{j\sigma'}c_{j-\sigma'}\delta_{ij}\delta_{\sigma\sigma'}\delta_{\sigma-\sigma'} \nonumber \\&& + c^{\dag}_{j\sigma'}c_{j\sigma'}c^{\dag}_{j-\sigma'}c_{j-\sigma'} c_{i\sigma}\big). \label{eq:A10}   
\end {eqnarray}    
We used $c_{i\sigma}c^{\dag}_{j\sigma'} = 1 - c^{\dag}_{j\sigma'}c_{i\sigma}$, $c_{i\sigma}c_{j\sigma'} = -c_{j\sigma'}c_{i\sigma}$, $n_{j\sigma'} = c^{\dag}_{j\sigma'}c_{j\sigma'}$ and $n_{j-\sigma'}c_{j\sigma'} = c_{j\sigma'}n_{j-\sigma'}$. Therefore, 
\begin {eqnarray}
&&\xi\big(c_{i\sigma}c^{\dag}_{j\sigma'}c_{j\sigma'}c^{\dag}_{j-\sigma'}c_{j-\sigma'} - c^{\dag}_{j\sigma'}c_{j\sigma'}c^{\dag}_{j-\sigma'}c_{j-\sigma'} c_{i\sigma}\big) \nonumber \\&& = \xi n_{i-\sigma}c_{i\sigma}. \label{eq:A11}   
\end {eqnarray}    
In the last step, we have replaced $j$, $-\sigma'$ and $\sigma'$ with $i$, $-\sigma$ and $\sigma$, accordingly via the Kronecker deltas and $n_{i\sigma}c_{i\sigma} = c^{\dag}_{i\sigma}c_{i\sigma}c_{i\sigma} = 0$. Applying the creation and annihilation operators to the left or the right will give two Kronecker deltas one for the site and one for the spin indices. We can now write
\begin {eqnarray}
&&i\hbar\frac{\partial\mathcal{G}(t-t')}{\partial t} \nonumber \\&& = 1 - \frac{i}{\hbar}\theta(t-t')\langle\{\big(E^{(0)}c_{i\sigma} + \xi n_{i-\sigma}c_{i\sigma}\big),c^{\dag}_{j\sigma}(t')\}\rangle   
\nonumber \\&& = 1 - E^{(0)}\frac{i}{\hbar}\theta(t-t')\langle\{c_{i\sigma}(t),c^{\dag}_{j\sigma}(t')\}\rangle - \nonumber \\&& \xi\frac{i}{\hbar}\theta(t-t')\langle\{n_{i-\sigma}c_{i\sigma}(t),c^{\dag}_{j\sigma}(t')\}\rangle. \label{eq:A12}   
\end {eqnarray}    
Using Eq.~(\ref{eq:A1})
\begin {eqnarray}
\bigg[i\hbar\frac{\partial}{\partial t} - E^{(0)}\bigg]\mathcal{G}(t-t') = 1 + \xi\mathcal{G}^{(1)}(t-t'). \label{eq:A13}   
\end {eqnarray}    
We define
\begin {eqnarray}
\mathcal{G}^{(1)}(t-t') = - \frac{i}{\hbar}\theta(t-t')\langle\{n_{i-\sigma}c_{i\sigma}(t),c^{\dag}_{j\sigma}(t')\}\rangle, \label{eq:A14}   
\end {eqnarray}    
\begin {eqnarray}
&&i\hbar\frac{\partial \mathcal{G}^{(1)}(t-t')}{\partial t} = \delta(t-t')\langle\{n_{i-\sigma}c_{i\sigma}(t),c^{\dag}_{j\sigma}(t')\}\rangle \nonumber \\&& - \frac{i}{\hbar}\theta(t-t')\bigg\langle\bigg\{n_{i-\sigma}i\hbar\frac{\partial c_{i\sigma}(t)}{\partial t},c^{\dag}_{j\sigma}(t')\bigg\}\bigg\rangle. \label{eq:A15}   
\end {eqnarray}    
We have dropped $\xi$ from Eq.~(\ref{eq:A15}) for the time being (because it is an eigenvalue), and we will tack $\xi$ back into the final equation after finding $\mathcal{G}^{(1)}(t-t')$. Now, using Eq.~(\ref{eq:A6}) 
\begin {eqnarray}
&&i\hbar\frac{\partial \mathcal{G}^{(1)}(t-t')}{\partial t} = \langle n_{i-\sigma}\rangle \nonumber \\&& - \frac{i}{\hbar}\theta(t-t')\bigg\langle\bigg\{[n_{i-\sigma}c_{i\sigma}(t),H^{\texttt{H}}_{\rm IET}],c^{\dag}_{j\sigma}(t')\bigg\}\bigg\rangle. \nonumber \\&& \label{eq:A16}   
\end {eqnarray}    
Next,
\begin {eqnarray}
&&[n_{i-\sigma}c_{i\sigma}(t),H^{\texttt{H}}_{\rm IET}] \nonumber \\&& = E^{(0)}\big(n_{i-\sigma}c_{i\sigma}c^{\dag}_{j\sigma'}c_{j\sigma'} - c^{\dag}_{j\sigma'}c_{j\sigma'}n_{i-\sigma}c_{i\sigma}\big) \nonumber \\&& + \xi\big(n_{i-\sigma}c_{i\sigma}c^{\dag}_{j\sigma'}c_{j\sigma'}c^{\dag}_{j-\sigma'}c_{j-\sigma'} \nonumber \\&& - c^{\dag}_{j\sigma'}c_{j\sigma'}c^{\dag}_{j-\sigma'}c_{j-\sigma'}n_{i-\sigma}c_{i\sigma}\big). \nonumber \\&& \label{eq:A17}   
\end {eqnarray}    
\begin {eqnarray}
&&E^{(0)}n_{i-\sigma}\bigg[c_{i\sigma}c^{\dag}_{j\sigma'}c_{j\sigma'} - c^{\dag}_{j\sigma'}c_{j\sigma'}c_{i\sigma}\bigg] = E^{(0)}n_{i-\sigma}c_{i\sigma}. \nonumber \\&& \label{eq:A18}   
\end {eqnarray}    
\begin {eqnarray}
&&\xi\bigg[n_{i-\sigma}c_{i\sigma}c^{\dag}_{j\sigma'}c_{j\sigma'}c^{\dag}_{j-\sigma'}c_{j-\sigma'} - c^{\dag}_{j\sigma'}c_{j\sigma'}c^{\dag}_{j-\sigma'}c_{j-\sigma'}n_{i-\sigma}c_{i\sigma}\bigg] \nonumber \\&& = \xi n_{i-\sigma}c_{i\sigma}, \label{eq:A19}   
\end {eqnarray}    
we used $\big(c_{i\sigma}\big)^2 = 0$ and 
\begin {eqnarray}
&& n_{i-\sigma}^2 = c^{\dag}_{i-\sigma}c_{i-\sigma}c^{\dag}_{i-\sigma}c_{i-\sigma} = c^{\dag}_{i-\sigma}\big(1 - c^{\dag}_{i-\sigma}c_{i-\sigma}\big)c_{i-\sigma} \nonumber \\&& = c^{\dag}_{i-\sigma}c_{i-\sigma} - c^{\dag}_{i-\sigma}c^{\dag}_{i-\sigma}c_{i-\sigma}c_{i-\sigma} = n_{i-\sigma}. \label{eq:A20}   
\end {eqnarray}    
Therefore
\begin {eqnarray}
&&[n_{i-\sigma}c_{i\sigma}(t),H^{\texttt{H}}_{\rm IET}] = \xi n_{i-\sigma}c_{i\sigma}(t) + E^{(0)}n_{i-\sigma}c_{i\sigma}(t), \nonumber \\&& \label{eq:A21}   
\end {eqnarray}    
using Eq.~(\ref{eq:A21}) and Eq.~(\ref{eq:A16}) 
\begin {eqnarray}
&&i\hbar\frac{\partial \mathcal{G}^{(1)}(t-t')}{\partial t} \nonumber \\&& = \langle n_{i-\sigma}\rangle - \xi \frac{i}{\hbar}\theta(t-t')\bigg\langle\bigg\{[n_{i-\sigma}c_{i\sigma}(t),c^{\dag}_{j\sigma}(t')\bigg\}\bigg\rangle \nonumber \\&& - E^{(0)} \frac{i}{\hbar}\theta(t-t')\bigg\langle\bigg\{[n_{i-\sigma}c_{i\sigma}(t),c^{\dag}_{j\sigma}(t')\bigg\}\bigg\rangle. \label{eq:A22}   
\end {eqnarray}
Using Eq.~(\ref{eq:A14})     
\begin {eqnarray}
&&i\hbar\frac{\partial \mathcal{G}^{(1)}(t-t')}{\partial t} \nonumber \\&& = \langle n_{i-\sigma}\rangle + \xi \mathcal{G}^{(1)}(t-t') + E^{(0)}\mathcal{G}^{(1)}(t-t'), \nonumber \\&& \bigg[i\hbar\frac{\partial}{\partial t} - E^{(0)} - \xi\bigg]\mathcal{G}^{(1)}(t-t') = \langle n_{i-\sigma}\rangle. \label{eq:A23}   
\end {eqnarray}
Therefore (after using $i\hbar\partial/\partial t \rightarrow E$)
\begin {eqnarray}
\mathcal{G}^{(1)}(t-t') = \frac{\langle n_{i-\sigma}\rangle}{E - E^{(0)} - \xi}. \label{eq:A24}   
\end {eqnarray}
Using Eq.~(\ref{eq:A24}) and Eq.~(\ref{eq:A13})
\begin {eqnarray}
\big(E - E^{(0)}\big)\mathcal{G}(t-t') = 1 + \frac{\xi\langle n_{i-\sigma}\rangle}{E - E^{(0)} - \xi}.
\end {eqnarray}
Consequently
\begin {eqnarray}
&&\mathcal{G}(t-t') = \frac{\langle n_{i-\sigma}\rangle}{E - (E^{(0)} + \xi) + i\delta} + \frac{1 - \langle n_{i-\sigma}\rangle}{E - E^{(0)} + i\delta}, \nonumber \\&& \label{eq:A25}   
\end {eqnarray}
after introducing the arbitrary convergence factor, $\delta$.


\end{document}